\documentclass[twocolumn]{aastex701}

\usepackage{graphicx}
\usepackage[suffix=]{epstopdf}
\usepackage{natbib}
\usepackage{amsmath}
\usepackage{url}
\usepackage{xspace}
\usepackage{listings}
\usepackage{hyperref}

\providecommand{\e}[1]{\ensuremath{\times 10^{#1}}}

\shorttitle{Technosignatures of ISOs}
\shortauthors{Davenport et al.}

\DeclareRobustCommand{\okina}{%
  \raisebox{\dimexpr\fontcharht\font`A-\height}{%
    \scalebox{0.8}{`}%
  }%
}

\newcommand{\linccfw}{LSST Interdisciplinary Network for Collaboration and Computing, Tucson, USA}
\newcommand{\diracuw}{Department of Astronomy and the DiRAC Institute, University of Washington, 3910 15th Avenue NE, Seattle, WA 98195, USA}
\newcommand{\nau}{Department of Astronomy and Planetary Science, Northern Arizona University, Flagstaff, USA}

\newcommand{\ucbastro}{Department of Astronomy, University of California, Berkeley, 501 Campbell Hall 3411, Berkeley, CA 94720, USA}
\newcommand{\bsrc}{Berkeley SETI Research Center, University of California, Berkeley, 501 Campbell Hall 3411, Berkeley, CA 94720, USA}
\newcommand{\seti}{SETI Institute, 339 Bernardo Ave, Suite 200, Mountain View, CA 94043, USA}
\newcommand{\blox}{Breakthrough Listen, University of Oxford, Department of Physics, Denys Wilkinson Building, Keble Road, Oxford, OX1 3RH, UK}

\newcommand{\PSUAA}{Department of Astronomy \& Astrophysics, 525 Davey Laboratory, 251 Pollock Road, Penn State, University Park, PA, 16802, USA}
\newcommand{\PSUCEHW}{Center for Exoplanets and Habitable Worlds, 525 Davey Laboratory, 251 Pollock Road, Penn State, University Park, PA, 16802, USA}

\newcommand{\PSETI}{Penn State Extraterrestrial Intelligence Center, 525 Davey Laboratory, 251 Pollock Road, Penn State, University Park, PA, 16802, USA}

\begin{document}

\title{Technosignature Searches of Interstellar Objects}

\author[0000-0002-0637-835X]{James R. A. Davenport}
\affiliation{\diracuw}
\email{jrad@uw.edu}

\author[0000-0001-7057-4999]{Sofia Z. Sheikh}
\affiliation{\seti}
\affiliation{\bsrc}
\email{ssheikh@seti.org}

\author[0000-0003-4823-129X]{Steve Croft}
\affiliation{\blox}
\affiliation{\seti}
\affiliation{\bsrc}
\email{scroft@berkeley.edu}

\author[0000-0003-1515-4857]{Brian C. Lacki}
\affiliation{\blox}
\email{astrobrianlacki@gmail.com}

\author[0000-0001-6160-5888]{Jason T.\ Wright}
\affiliation{\PSUAA}
\affiliation{\PSUCEHW}
\affiliation{\PSETI}
\email{astrowright@gmail.com}

\author[0000-0001-5578-359X]{Chris Lintott}
\affiliation{Department of Physics, University of Oxford, Denys Wilkinson Building, Keble Road, Oxford, OX1 3RH}
\email{chris.lintott@physics.ox.ac.uk}

\author[0000-0002-4948-7820]{Adam Frank}
\affiliation{Department of Physics and Astronomy, University of Rochester, Rochester New York, 14627-0171}
\email{Afrank@pas.rochester.edu}

\author{T.~Joseph~W.~Lazio}
\affiliation{Jet Propulsion Laboratory, California Institute of Technology, 4800 Oak Grove Dr, Pasadena, CA  91109, USA}
\email{Joseph.Lazio@jpl.caltech.edu}

\author[0000-0001-7335-1715]{Colin Orion Chandler}
\affiliation{\linccfw}
\affiliation{\diracuw}
\affiliation{\nau}
\email{coc123@uw.edu}

\author[0000-0003-2828-7720]{Andrew P.\ V.\ Siemion}
\affiliation{\blox}
\affiliation{\seti}
\affiliation{\bsrc}
\email{andrew.siemion@physics.ox.ac.uk}

\author[0000-0003-2633-2196]{Stephen DiKerby}
\affiliation{Department of Physics and Astronomy \\Michigan State University, East Lansing, MI 48820, USA}
\email{dikerbys@msu.edu}

\author[0000-0002-9112-1734]{Ellie White}
\affiliation{Department of Mathematics and Physics, Marshall University, Huntington WV 25755, USA}
\email{elliewhite1420@gmail.com}

\author[0009-0000-2879-6539]{Valeria Garcia Lopez}
\affiliation{Department of Physics, Furman University, Greenville, South Carolina 29613, USA}
\email{valeriagl8208@gmail.com}

\author[0009-0000-3814-100X]{Emma E.\ Yu}
\affiliation{\ucbastro}
\email{emmaeyu1@gmail.com}

\author[0000-0002-8069-3139]{Maxwell K. Frissell}
\email{thefrissmax@gmail.com}
\affiliation{\diracuw}
\affiliation{\nau}

\author[0009-0007-3755-0021]{Naomi Morato} 
\email{nmorato@uw.edu}
\affiliation{\diracuw}

\author[0009-0004-0059-3229]{Devanshi Singh}
\email{ds2004@uw.edu}
\affiliation{\diracuw}

\author[0009-0004-0944-9098]{Jinshuo Zhang}
\email{jinshuoz@uw.edu}
\affiliation{\diracuw}

\author[0000-0003-2874-6464]{Peter Yoachim}
\affiliation{\diracuw}
\email{yoachim@uw.edu}

\author[0000-0002-0726-6480]{Darryl Z. Seligman}
\altaffiliation{NSF Astronomy and Astrophysics Postdoctoral Fellow}
\affiliation{Department of Physics and Astronomy, Michigan State University, East Lansing, MI 48824, USA}
\email{dzs@msu.edu}

\begin{abstract}
With the discovery of the third confirmed interstellar object (ISO), 3I/ATLAS, we have entered a new phase in the exploration of these long-predicted objects. Though confirmed discovery of ISOs is quite recent, their utility as targets in the search for technosignatures (historically known as the Search for Extraterrestrial Intelligence -- SETI) has been discussed for many decades. With the upcoming NSF-DOE Vera C. Rubin Observatory's Legacy Survey of Space and Time (LSST), the discovery and tracking of such objects is expected to become routine, and thus so must our examination of these objects for possible technosignatures. Here we review the literature surrounding ISOs as targets for technosignatures, which provides a well-developed motivation for such exploration. We outline four broad classes of technosignatures that are well suited for ISO follow-up, including the type of data needed and the best timing for study. Given the limitations in the current understanding of ISOs, we show that care must be taken in identifying technosignatures based primarily on comparison to objects in the Solar System. We therefore provide a roadmap for careful and consistent study of the population of ISOs in the hope of identifying technosignatures.
\end{abstract}

\section{Introduction}

The search for evidence of technology of non-human origins, known colloquially as the Search for Extraterrestrial Intelligence (SETI), is a central pillar of astrobiology, aiming to constrain the presence of advanced life beyond Earth \citep{tarter2001}. While radio searches for technosignatures have historically dominated public and scientific attention, the field now encompasses a growing range of observational strategies and targets. Recent reviews have attempted to systematize this diversity, such as the ``Nine Axes of Merit'' framework proposed by \citet{sheikh2020}. This framework covers traditional direct electromagnetic signals, physical technosignatures such as artifacts or engineered structures, as well as incidental technosignatures like atmospheric pollutants. 

A wide variety of possible physical technosignatures in the Solar System have long been of interest \citep{bracewell1960}. 
Such technosignatures could arise from either active probes or passive or defunct objects, which could be unique moving objects, or embedded on planetary or minor-body surfaces \citep{tarter2001,sheikh2020,KISS_DataSETI}. 
To date only a small portion of the Solar System has been systematically searched for technosignatures, primarily planetary surfaces, the Moon, and within stable orbits \citep{davies2013,wright2018a,lesnikowski2024}. 
Importantly, solar system technosignatures could imply that advanced life is long-lived enough to become mobile on Galactic scales. They could also imply that such visits occurred deep in the Solar System's past, which would be relevant to addressing the ``Fermi Paradox'' \citep{Carroll-Nellenback2019,SchmidtFrank2019}.

Interstellar objects (ISOs) have gained broad attention within both the planetary science and astrophysics communities, as small bodies likely ejected from other planetary systems. As such, ISOs present a new means of placing our solar system in the larger context of planetary formation and evolution.
The first confirmed \hbox{ISO}, 1I/\okina Oumuamua, was discovered in~2017 \citep{Williams17} after a perihelion passage near the Sun of $\sim0.25$\,AU on a hyperbolic orbit. Though small in size (with radius $\sim100$\,m), it generated enormous interest due to its unusual trajectory, lack of observable out-gassing tail, elongated shape  \citep{mashchenko2019, taylor2023} inferred from dramatic brightness variations \citep{meech2017, jewitt2017, knight2017, bannister2017, fitzsimmons2018, drahus2018}, non-gravitational acceleration 
\citep{micheli2018}, and overall novelty.

As with ISOs that arise naturally as small bodies ejected from other planetary systems, extraterrestrial interstellar probes would also enter the Solar System on hyperbolic orbits. 
The notion of physical artifacts from another civilization is plausible based on the Pioneer~10 and~11, Voyager~1 and~2, and New Horizons spacecraft, all of which either have left the Solar System or are on escape trajectories.
Further, there have been multiple concept studies for missions designed to leave the Solar System, with the most recent being the Interstellar Probe mission concept \citep{interstellarprobe}.

None of these human-made spacecraft are designed to be operational on the time scales relevant for reaching a nearby star ($> 10\,000$\,yr).
Nonetheless, these spacecraft establish an existence proof that it is feasible to launch interstellar probes that can reach other stars on timescales much shorter than the age of the Galaxy, without resorting to any unknown or speculative physics.
Whether operational or not, and whether aimed intentionally at the Solar System or not, finding a spacecraft or other artifact of non-terrestrial origin in the Solar System would be an unambiguous technosignature.

Shortly after the discovery of 1I/\okina Oumuamua,
a number of facilities conducted follow-up observations to search for possible technosignatures. These included the Allen Telescope Array \citep[\hbox{ATA},][]{harp2019}, Breakthrough Listen with the Green Bank Telescope \citep[\hbox{GBT},][]{enriquez2018}, and the Murchison Widefield Array \citep{tingay2018a}.

The suggestion that 1I/\okina Oumuamua might be of technological origin was notably advanced by \citet{bialy2018} and \citet{loeb2022}. This hypothesis has not gained support from other studies \citep{curran2021}, with debate centered primarily around possible non-gravitational acceleration it exhibited as it exited the Solar System \citep{micheli2018}. For example, \citet{bialy2018} argued that 1I/\okina Oumuamua's non-gravitational acceleration could be caused if it had a millimeter or thinner axis, perhaps with an artificial provenance such as something similar to a solar sail. However, \cite{Zhou2022} demonstrated that such a morphology did not provide an adequate fit to the photometric brightness variations, primarily because of the non-zero flux minimum exhibited every $\sim8$ hours \citep[e.g.][]{Belton2018,jewitt2023,wright2023medium}. 
The non-gravitational acceleration exhibited by 1I/\okina Oumuamua can be explained by out-gassing or radiation pressure from a natural object \citep{Levine2021_h2,Bergner2023,jackson20211i,desch20211i,Seligman2020,fuglistaler2018,MoroMartin2019,Luu2020}. 
Indeed, the consensus view is that technosignature assertions about 1I/\okina Oumuamua are unjustified given our current knowledge \citep{oumuamua-issi2019}. 
The wealth of observations and analysis for 1I/\okina Oumuamua to constrain possible technosignatures did improve the understanding of the object, and have helped motivate future work in this area.

The second confirmed ISO, 2I/Borisov, was discovered in 2019 \citep{borisov_2I_cbet}. Unlike 1I/\okina Oumuamua, 2I/Borisov had clear cometary activity upon discovery, and was classified as an interstellar comet \citep{opitom2019,Jewitt2019,Fitzsimmons2019,Kareta2019,Xing2020,Bodewits2020,Cordiner2020}. This object also received technosignature follow-up from Breakthrough Listen radio observations \citep{lacki2021}.

Now 3I/ATLAS has been identified as a third interstellar object \citep{Denneau2025,seligman2025}. Multiple follow-up observations have indicated that 3I/ATLAS is an active cometary body, consistently identifying a faint coma and tail \citep[e.g.][]{chandler2025,Jewitt2025,Alarcon2025,Opitom2025,Puzia2025,Belyakov2025,Kareta2025,Jewitt2025_HST}, and with possible detections of periodic brightness variations \citep{Marcos2025,Santana-Ros2025}. Furthermore, spectroscopic follow-up indicates the presence of water ice in the coma \citep{yang2025} and water production via OH \citep{Xing2025}. Radio technosignature observations have also been conducted with the Allen Telescope Array, covering GHz- wide frequency ranges with high spectral resolution to search for narrow-band signals that could indicate artificial origin (Sheikh et al {\it in prep}). Speculation about possible artificial origins, and how its orbit may evolve over time, have also been made \citep{hibberd2025}.  The rapidly growing suite of observations for 3I/ATLAS strongly supports the conclusion that it is a comet. To date there is no credible evidence that any of the three ISOs detected so far are anything other than natural objects.

Given the wide interest in detecting and characterizing the population of ISOs expected in the next decade, our goal is to motivate the systematic study of these objects for possible technosignatures. 
In \S\ref{sec:history} we give a historically driven overview of ISOs as technosignature targets, which provides clear motivation for many ongoing technosignature searches. We outline the classes of possible technosignatures that could be detected for ISOs, and their recommended observing strategies, in \S\ref{sec:search}. Importantly, many of the observations needed for such constraints are highly complementary to other solar system science cases, and could be applied also to studies of asteroids and comets. In \S\ref{sec:summary} we summarize our recommendations for technosignature searches from ISOs, and provide a brief concluding discussion in \S\ref{sec:dicussion}.

\section{Historic Motivation of Interstellar Objects as Technosignature Targets} 
\label{sec:history}

The notion of advancing SETI through the study of anomalous solar system bodies, particularly ISOs, has existed for many decades, but until recently, there haven't been any targets for study. The concept was popularized in fiction by Arthur C. Clarke in \textit{Rendezvous with Rama} (1973), where a large object is detected passing through the Solar System that is found to have a technological origin through its cylindrical shape. 
Despite no strong evidence for technological origins or anomalies from the three known ISOs at present, the continued search for technosignatures from 3I/ATLAS and future ISOs is supported by numerous studies on the possible scenarios and signals that such interstellar technology may exhibit.

Since humanity has sent many of our own spacecraft on interstellar trajectories, we know that it is possible to send probes to other stars. \citet{bracewell1960}  discussed the possibility of interstellar communication via physical probes, which would slowly travel the Galaxy trying to make their presence known. 
\citet{von-neumann1966} developed the first mathematical models of self-replicating machines, which \citet{tipler1980} and \citet{freitas1985} used as a basis for advancing the ``Bracewell Probe'' scenario to more efficiently spread throughout the Galaxy. Such ``Von Neumann Probes'' would need raw materials to carry out self replication, such as from the Oort Cloud or Jupiter Trojan system.

Many different forms of technosignatures could arrive as an ISO \citep{gertz2021}. Technology could either be active throughout an object's interstellar journey, or wake from a dormant state upon arrival in the Solar System (conceptually similar to ``lurkers'' stationed in the Solar System that have not made contact yet). Defunct technology may also be found, such as spacecraft whose power supplies have long since become inactive. For example, the Voyager spacecraft power supplies are expected to become inactive sometime in the next decade \citep{winter2000}, long before they approach any nearby star \citep{bailer-jones2019}.
ISO technosignatures could be in the form of standalone spacecraft, or embedded with natural objects such as comets that are on interstellar trajectories. Technology could even be buried under the surface of such objects \citep{freitas1980}, and which may be revealed after material sublimates away when the ISO approaches the Sun.

ISOs could also facilitate the physical migration or transport of life. 
Besides built spacecraft or ``World Ships'' \citep{hein2012}, these objects might include structures made from modified, natural ISOs, such as proposals to convert asteroids into habitats \citep{oneill1974,Miklavcivc2022}. 
\citet{finney1985}, for example, considered the idea of ``Interstellar Nomads'', who would use generational habitats made using comets that would cross interstellar space over $\sim$10$^5$ year timescales. Such slow travel between stars could make efficient use of resources found along the way \citep{gilster2013}.
While these timescales are long compared with human lifetimes, they are extremely short compared to stellar or galactic evolution \citep{zuckerman1985}.

Given the extreme distance between star systems, sending physical objects between stars may be advantageous for an extraterrestrial civilization, compared with direct communication. Small probes can be launched at high speeds that make scientific exploration between stars feasible, such as the proposed Breakthrough Starshot model for launching ultra-lightweight probes to nearby stars \citep{parkin2018}. Probes are able to carry a large volume of information via physical media, which can be more efficient (i.e. higher bandwidth) than direct radio communication \cite[e.g. see][]{hippke2017,kerby2021}.

Searching for extraterrestrial technology within our solar system is also highly advantageous in the search for technosignatures \citep{freitas1983b}. As \citet{gertz2016} notes, complete monitoring of the entire Solar System is likely more tractable than continuous observation of millions to billions of stellar targets for technosignatures. The millions of known minor bodies in the Solar System can be tracked, and newfound or anomalous objects can be identified through repeated wide field observations. Beyond ISOs on hyperbolic trajectories, probes could be located as outliers among a variety of populations across the Solar System, such as the asteroid belt \citep{papagiannis1978}, planetary Lagrange points \citep{freitas1980, valdes1983}, and even as Earth co-orbiters \citep{benford2019}. Note that while we are focused on ISOs in this manuscript, much of the history, observation, and technosignature strategies discussed are applicable to studies of bound solar system objects, including moons, comets, and asteroids \citep{chapman-rietschi1994,arkhipov1995,haqq-misra2012}.

Extraterrestrial probes in the Solar System can be potentially examined in great detail. Since they can be studied over many nights, a wide variety of observations can be gathered, and strict constraints can be placed on their properties and behavior. Given their relatively close proximity, such interstellar technology requires extremely low amounts of power to produce detectable emission compared with interstellar communication, and such transmissions could be easily detected. Strong upper limits on the presence of such interstellar technosignatures can also be generated through detailed observation and modeling of the Solar System and discovered ISO populations, and using present-day sky surveys such as the NSF-DOE Vera C.~Rubin Observatory \citep{ivezic2019}.

\section{Technosignature Search Strategies}
\label{sec:search}

ISOs are a new and relatively unexplored population of objects, originating from distant and unknown star systems. 
With a sample of only three ISOs known to date, it is extremely difficult to accurately constrain what the true distributions of natural or expected properties are. 
As with all SETI approaches, technosignature studies from ISOs must compete with the complex behavior of natural minor bodies. 
Comparison with the small bodies of our Solar System is useful, but given their novel origins we might yet find differences between our own Solar System’s contents and that of the wider ISO population.

Each solar system is predicted to eject an enormous number of objects into the interstellar medium, and we expect a large number of natural bodies like comets and asteroids to appear as ISOs in our sky surveys \citep{moro-martin2009, do2018, jewitt2023}. Numerous studies have simulated the population of natural ISOs that will be detectable, and evaluated the expected detection efficiency for upcoming surveys \citep{Cook2016,Engelhardt2014,Hoover2022,Marceta2023a,Marceta2023b}. The Rubin Observatory alone is expected to observe between 6 and 51 ISOs over the next decade \citep{dorsey2025}, though this value is highly sensitive to assumed number density and size-frequency-distribution of the population.
Note that while such population synthesis models assume ISOs are dynamically heated via similar mechanisms that govern the kinematics of stars, probes may not obey the same kinematic distributions. As with all technosignature studies, discovery of artificial ISOs will likely be biased towards objects or probes that are intended for detection (i.e. are ``loud''). We are potentially sensitive to e.g. derelict probes that are ``quiet'', if they exhibit sufficiently anomalous behavior.

We propose four categories of ISO properties that should be studied to constrain the likelihood that any given ISO is or is not technological. Besides classical SETI approaches for directed transmissions (\S \ref{sec:radio}), these categories are primarily focused on identifying outliers in the main observables for minor bodies (e.g., their kinematics and surface properties).

\subsection{Anomalous Trajectories}
\label{sec:trajectory}

One of the most powerful technosignature constraints on ISOs comes from their kinematics or motion through the Solar System. 
If a technological object were to suddenly engage an engine or thruster system, the ISO could appear to undergo a rapid non-gravitational acceleration.

There is some precedent for such an event.
During a radar track of the near-Earth asteroid 2001~DO${}_{47}$ by the Goldstone Solar System Radar (GSSR), dramatic changes were observed in the radar return and inferred orbital properties of the object \citep[J.~Giorgini, private communication;][]{2001IAUC.7589....3G}.
Subsequent investigation revealed that the object was in fact the WIND spacecraft, which had been classified mistakenly as a near-Earth asteroid, and that it had conducted a trajectory course maneuver during the GSSR observation.

In the most extreme scenario, a change in the velocity of an ISO could be so great it would cause existing algorithms for ``linking'' moving object detections between nights to fail \citep{holman2018,Taylor2024}. The ISO would seem to ``disappear'' from subsequent telescope observations at the predicted position, and a spurious new object would be detected. Any rapid changes in acceleration, particularly without any notable increase in cometary activity, should be treated as a high priority anomaly and trigger immediate and coordinated follow-up observations.

\begin{figure*}[!ht]
\centering
\includegraphics[width=6in]{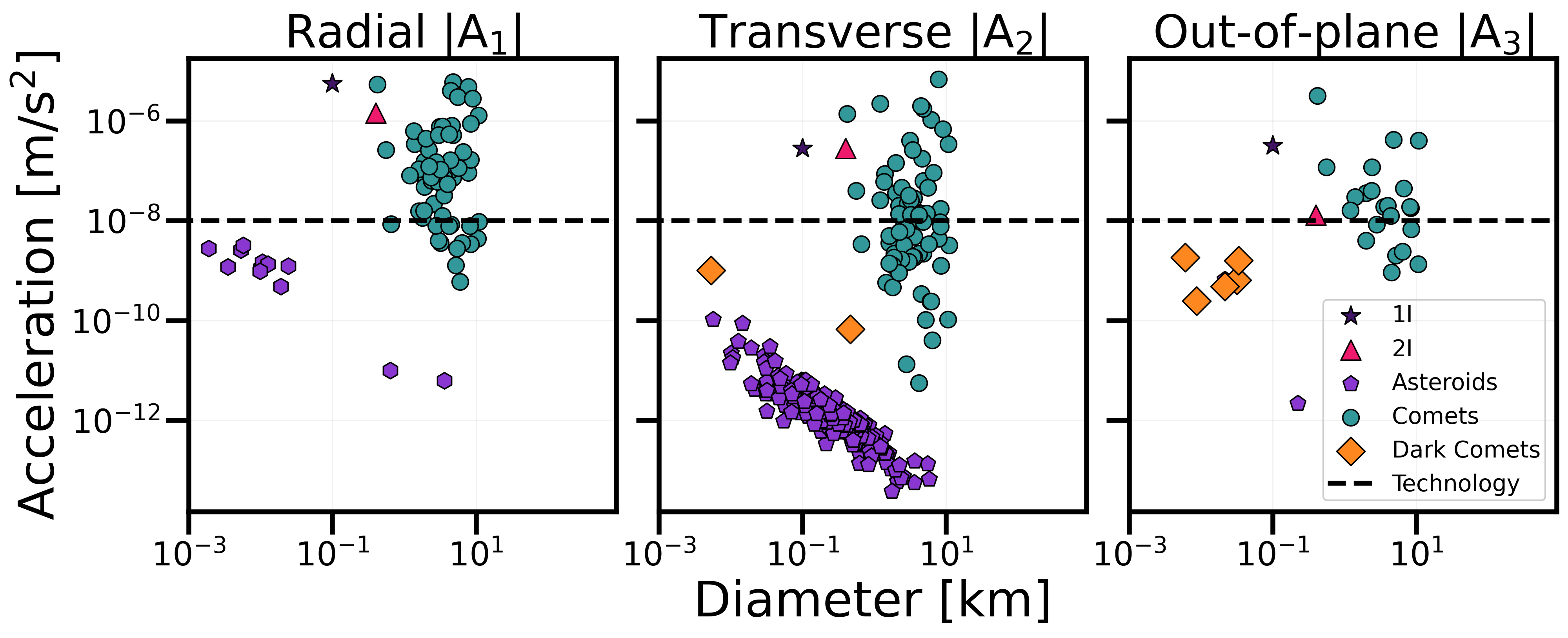}
\caption{Non-gravitational accelerations identified in natural asteroids and comets as a function of object size. An approximate limit for technology that has a low mass-to-area ratio (i.e. strongly accelerated by radiation pressure) is shown (dashed line). 
Adapted from \citet{Seligman2023}.
}
\label{fig:nongravs}
\end{figure*}

Wide field surveys, particularly the Rubin Observatory in the southern hemisphere, will be the primary means for early detection, as well as continuous surveillance of ISO positions. In order to ensure the most robust constraints on anomalous non-gravitational accelerations, these surveys must maintain sub-arcsecond accuracy and precision in their astrometry. 
Constant monitoring of ISOs, at least nightly if possible, ensures that any extreme accelerations are hopefully tracked and the ISO does not ``disappear.''
Dedicated imaging from both small and large aperture facilities is useful, particularly as the ISO passes through dense (e.g., low Galactic latitude) fields of background objects that impair precise centroiding. Large aperture telescopes should continue to monitor the ISO’s astrometry as the object fades to provide the longest lever arm on the dynamics possible.

More strategic accelerations may also be evident from technological activity. These could include ``gravity assist'' maneuvers, orbital transfers, or braking. Such activity would likely occur at close approaches to planetary bodies or the Sun. If the ISO was intent on parking in the Solar System (i.e entering a stable and bound orbit), a very gradual deceleration may be apparent, which would stand out as increasingly anomalous as the object approached the Sun. For example, since 3I/ATLAS is moving at over 60 km s$^{-1}$, it would need a very powerful deceleration to reach a bound orbit within the inner solar system. 
If the ISO was instead using the Sun as a gravity assist towards another target, we would expect accelerations near perihelion, and that the outbound deceleration as it left the Solar System would be discrepant from our models. Careful constraints of the inbound orbit, particularly in deep archival images (``precovery'' detections), will help firmly establish the inbound velocity of the ISO, which is critical for such anomaly detections. ISOs that are expected to pass extremely close to planetary bodies should be monitored constantly around such passages for signs of gravity-assist acceleration.

It is important to note that natural objects, including asteroids and comets, experience a suite of non-gravitational accelerations. For example, in Figure \ref{fig:nongravs} we show non-gravitational acceleration magnitude and components for a range of interstellar objects, asteroids, comets, and dark comets. These include accelerations from sublimation or cometary activity, radiation pressure \citep{Vokrouhlicky2000}, the Yarkovsky effect  \citep{Vokrouhlicky2015_ast4}, and the YORP effect. Hundreds of near-Earth objects have measured accelerations from the Yarkovsky effect \citep{Farnocchia2013,Greenberg2020} and a small number of asteroids have the same from radiation pressure \citep{Micheli2012,Micheli2013,Micheli2014,Mommert2014bd,Mommert2014md,Farnocchia2017TC25,Fedorets2020}.  Since these broadly involve solar heating or insolation, these non-gravitational accelerations will increase as the ISO approaches the Sun. Given the recent discovery of ISOs, as well as many possible ``dark comets'' \citep{seligman2024,Seligman2023b,Farnocchia2023} and ``active asteroids'' \citep[e.g.][]{chandler2024,2024come.book..767J}, the boundary between natural and conspicuous accelerations is currently under-constrained, and deserves robust theoretical exploration. As such, small amplitude accelerations alone should be treated with great care. The ISO trajectory should be compared constantly with industry-standard dynamics models, such as those used by the Minor Planet Center and JPL Horizons, rather than simple N-body or Keplerian approximations.

Other anomalies from the trajectories of ISOs are also worth searching for. Natural ISOs are expected to broadly follow the Galactic stellar kinematic distribution \citep[e.g. see Fig. 1 from ][]{hopkins2025}, and objects with highly discrepant velocities from this phase space distribution would be noteworthy. 
Further, the distribution of natural ISO origins is expected to be smooth (though not uniform) across the sky \citep{dorsey2025,hopkins2025,Marceta2023a,Marceta2023b}. Given the uncertainties in stellar kinematics alone, it is not possible at present to identify the specific stellar system that an ISO originates from.
Multiple ISOs coming from a very focused radiant or direction of origin, or which are clustered in velocity space, and particularly if they are discovered close in time, could be an indication of a probe system, rather than the expected scattering of proto-planetary material across the Galaxy.

\subsection{Anomalous Spectra or Colors}\label{sec:spectra}

Along with orbital dynamics and object kinematics, broadband optical colors can be used to detect anomalous objects across the Solar System \citep{rogers2024}.
Eccentricity and colors, for example, can be used to identify asteroid families due to varying object compositions \citep[e.g.][]{parker2008}. 
Extreme outliers in photometric color space, particularly in many-color spaces from e.g. Rubin, could indicate unnatural surface materials. This could be due to surface paint or coatings, or materials with unusual reflectance properties like glass or metal. Such materials might also create unusual polarization signatures or glints \citep{lacki2019,villarroel2022}. Specular reflections have been proposed as a source for technosignatures, and could be especially noteworthy as short-timescale transients in the light curve \citep{lacki2019}. Getting multi-band photometry, particularly at early detection before the ISO is strongly heated by the Sun, is critical for quickly and efficiently classifying its color.

\begin{figure}[]
\centering
\includegraphics[width=3.25in]{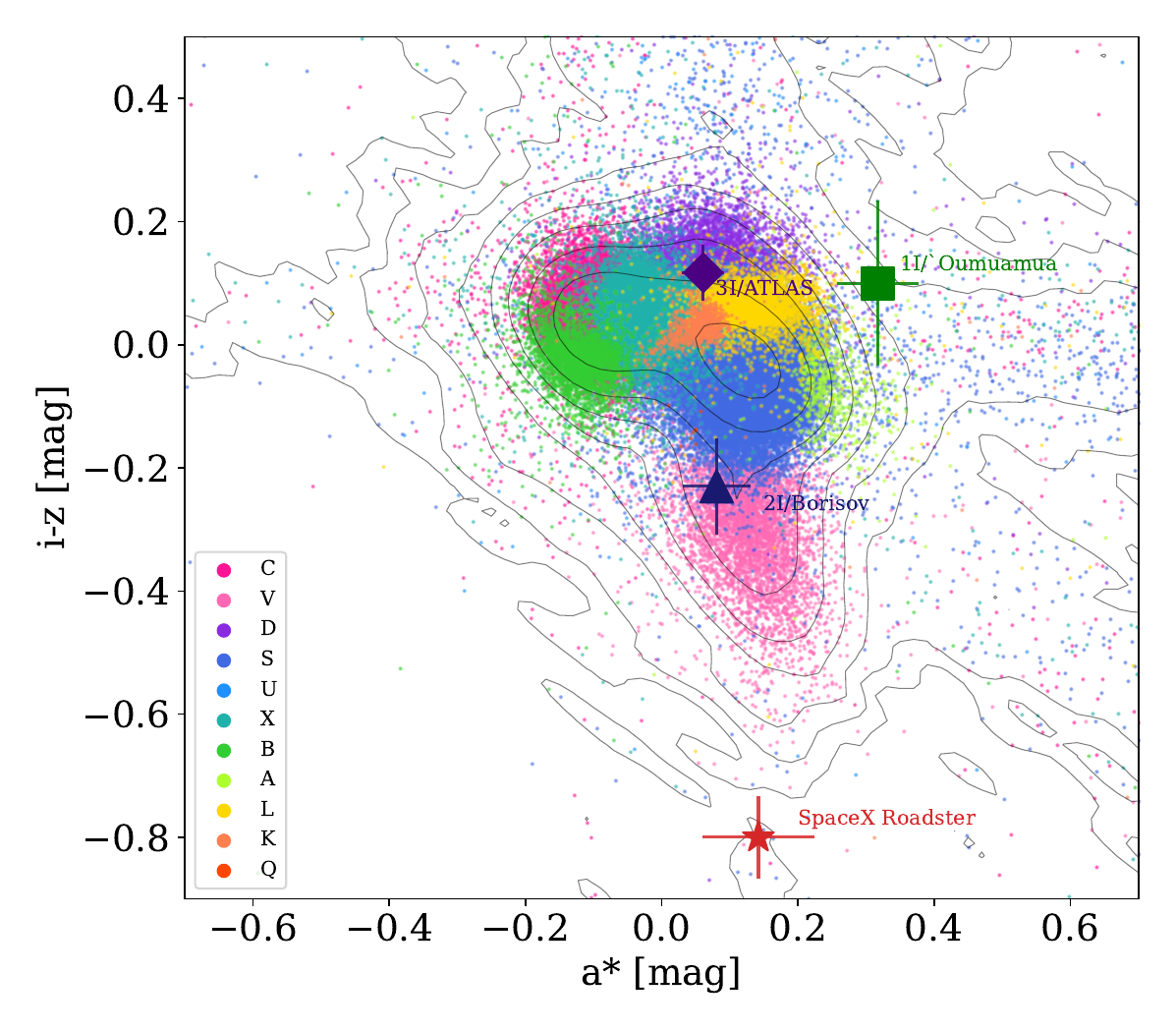}
\caption{Color-color plot for asteroids from \citet{sergeyev2021}, based on the parameter space defined by \citet{ivezic2001} to identify asteroid families and surface compositions. The most likely asteroid family membership for each object determined by \citet{sergeyev2021} is noted by point color, and the figure legend. Contours show a kernel density estimator of the density of all 90,893 asteroids in this sample in this space. 
The three known ISOs show different colors, but are all consistent with populations of known solar system objects.
We also show the measured color of the SpaceX Roadster from DES.
}
\label{fig:tesla}
\end{figure}

While asteroid colors are instrumental in our understanding of their composition and history of the Solar System \citep{demeo2014}, the color of ISOs alone are likely not a definitive technosignature. 
For example, outer solar system objects are known to be very red, and show a diversity of colors \citep{sheppard2010}. A wide range of photometric colors have been observed for many asteroid populations \citep{hainaut2012,wong2017}.
In Figure \ref{fig:tesla} we show the color--color space for asteroid studies defined by \citet{ivezic2001}, using the benchmark catalog of asteroids from \citep{sergeyev2021}. Here we select 90,893 objects with photometric uncertainties in $griz$ bands less than 0.05 mag.
The published colors from $griz$ photometry for the three known ISOs are included in Figure \ref{fig:tesla}. We adopt the colors of 1I/\okina Oumuamua from  \citep{meech2017}, 2I/Borisov from \citep{bolin2020}, and 3I/ATLAS from \citep{beniyama2025}. While their colors are distinct from each other, the ISOs so far have colors consistent with natural objects. 
For an ISO to be an extreme outlier in photometric color, it likely would need a highly unusual surface material, such as from an anti-reflective coating \citep[e.g.][]{zinovev2023,poitras2025}, which could cause it to be effectively opaque in one filter and reflective in another.

We also show in Figure \ref{fig:tesla} the color--color location of the Tesla Roadster (a.k.a. ``Starman''), which was launched in 2018 and observed with the Dark Energy Survey \citep[DES][]{DES_dr2}, and was noted as having unique colors by \citet{karouzos2018}. 
For use in future studies, we list here the colors of the Tesla Roadster from DES provided by David Gerdes (Private Communication):
$g-r = 0.80 \pm 0.09$, 
$r-i = 0.00\pm 0.04$, 
$i-z -0.80 \pm 0.07$,
and $a* = 0.14 \pm0.08$.
While this human-made technosignature has colors that are not typically observed, there are multiple asteroids known with similar colors.

If the ISO is artificially illuminated, e.g., from lighting systems, then the brightness of the object will show an anomalous change in brightness over time. As \citet{loeb2012} shows, such technosignatures will follow a $r^{-2}$ change in brightness, rather than the $r^{-4}$ behavior expected from reflected light as a function of distance ($r$). Since cometary activity can substantially change the reflectance of an object, this technosignature will be most effectively searched when the object is far away, and particularly when inbound before it is heated by the Sun.

If the ISO has active technology, it will likely produce waste heat that could be detectable as infrared emission \citep{papagiannis1985}. As in the search for Dyson Spheres around distant stars, such waste heat is likely an unintentional technosignature from the use of extraterrestrial technology \citep{wright2014, sheikh2020}. Infrared imaging, particularly at mid-infrared bands such as from the WISE  \citep{wise} and upcoming NEOSurveyor missions \citep{neosurveyor}, and before the ISO becomes heated by its closest approach to the Sun or large bodies like planets, would be useful for constraining such emission. 
Repeated infrared imaging would constrain waste heat that begins to emit after a probe ``wakes up'' from a dormant state during an interstellar cruise.

Spectroscopy is especially useful for constraining anomalies in surface color, emission, or solar reflectance features, and for studying the composition of any cometary emission \citep[e.g.][]{swings1942,cambianica2021}. Low resolution ($R\sim1000$) spectroscopy can often be obtained in a few hours of observation with moderate aperture facilities, making it especially useful when the ISO is faint. 
This can be the most reliable way of measuring the object's colors, and looking for early signs of sublimation. High resolution spectroscopy is critical for precisely studying the composition of any ejected material. As with photometric outliers, anomalous features in spectroscopy may point to artificial coatings of the ISO, or even directed emission (which we discuss further in \S3.4). 

Objects that rapidly change their color or apparent spectral energy distribution are especially noteworthy. This could be due to intentional changes or modifications of the ISO surface, artificial material previously hidden being revealed due to sublimation, or the ignition of e.g. an ion thruster \citep{domonkos2000}.
Regular spectroscopic monitoring is recommended to ensure the ISO does not show transient emission features or changes in spectral profile.

Though it is beyond the capability of most current optical survey telescopes, anomalies could also be searched for in the polarization of an ISO compared with normal asteroids and comets. Polarization has been used to constrain the homogeneity of asteroid surfaces \citep{kwon2023}. Polarimetric measurements were obtained from 2I/Borisov \citep{Bagnulo2021}, which displayed an unusually high degree of polarization compared with typical comets. This was attributed to 2I/Borisov being relatively more pristine than solar system comets  \citep{Halder2023}. 
Objects with large variations in their polarization as a function of rotation or orbital phase, or that show rapid changes over time, would be especially interesting.

\subsection{Anomalous Shapes}
\label{sec:shape}

Like human made spacecraft and satellites, an interstellar probe might have a shape or profile that is anomalous compared with natural objects. This could include very thin profiles from e.g. solar sails,  geometric shapes such as cylinders that are not found in natural asteroids, or objects with complex modular structures. 
While observations of unnatural shapes could yield a fairly unambiguous technosignature detection, such measurements are extremely difficult to make with sufficient resolution, and likely impossible for most ISOs.

The shape or axis ratio of asteroids can be inferred through the modulation of reflected light as they rotate or tumble relative to our line of sight. Determining the true size of asteroids is intrinsically degenerate with the constraint on their albedo. Most asteroids, particularly smaller objects, show elongated or triaxial shape profiles \citep{cibulkova2016}. If the body is not spherical (or homogeneous in albedo), the amount of reflected light is modulated with the rotation angle. Since many objects are known to tumble or rotate about multiple axes, such as 1I/\okina Oumuamua, the observed flux modulation may be quasi-periodic. This complex behavior can be measured through flexible period finding algorithms such as Gaussian Process regression \citep{willecke-lindberg2021}. Since small body rotation periods -- including asteroids and ISOs -- can often be less than two days, high cadence imaging should be gathered for many successive nights, or using telescope networks like the Las Cumbres Observatory Global Telescope \citep[LCOGT][]{brown2013} or long-stare systems like the Transiting Exoplanet Survey Satellite \citep[TESS][]{tess} to estimate the rotation rates \citep{feinstein2025,Martinez-Palomera2025}. 
ISOs whose rotation profiles, or rotation rates, change significantly or suddenly are of particular interest, as they may indicate intentional control of the object. Note, however, that comets are known to experience rotation torques and even fragmentation due to sublimation as was observed with 2I/Borisov  \citep{jewitt2020,Kim2020}.  

The lack of any observed rotation period is also a powerful constraint on the nature of any ISO. Constant brightness implies either a substantial cometary coma (i.e. obscuring the rotating nucleus), a spherical and homogeneous body, or an object under active control. 
Probes that exhibit stable rotation (i.e. not tumbling about multiple axes) aligned with their motion vectors may be using centrifugal force to generate artificial gravity. ISOs should repeatedly have their rotation profiles measured throughout their solar system passage to constrain these scenarios.

Even if objects don’t show any significant rotation, either because their rotation axis is not favorably aligned, or because they have near spherical and homogeneous shapes, the change in illumination as the object orbits the Sun (the phase curve profile) can also help determine their size and shape.
Anomalous phase curve profiles may indicate artificial object shapes \citep{Zhou2022}. However, these profiles would have to be extreme outliers to be especially noteworthy, considering the wide variety of shapes seen across various asteroid families \citep{szabo2008}.
Monitoring the object's brightness throughout its solar system journey is essential for measuring the phase curve and estimating the absolute object size \citep{mahlke2021,martikainen2021}. 
Since inbound ISOs may not be uniformly distributed across the sky \citep{hopkins2025}, Northern hemisphere wide-field telescopes are especially needed to detect and monitor these objects, and constrain their phase curve profiles, particularly before they become heated by close passage of the Sun.

Direct, resolved imaging is also possible for some solar system bodies, if the object is very large or extremely close \citep{storrs1999,vernazza2021}.
However, the population of ISOs found so far does not appear to be dominated by 1,000 km bodies \citep{oumuamua-issi2019,jewitt2023}, and the majority are not expected to pass close enough to the Earth to image with existing facilities \citep{dorsey2025}.
As such, directly resolving the object is unlikely at present unless it comes precariously close to the Earth. Alternatively, an \textit{in situ} spacecraft mission to an interstellar object could provide resolved images \citep{Hein2019, Garber2022, Miller2022,hibberd2025,Yaginuma2025,Loeb2025b,Sanchez_2025,Seligman2018, Moore2021,Stern2024,Mages2022, Donitz2023,Landau2023}, such as the proposed Comet Interceptor mission \citep{Snodgrass2019, Jones2024} 

For asteroids and comets that pass very near the Earth ($<<$1\,AU), radar imaging is possible. At present, 138 Main-belt asteroids and 23 comets, as well as over 1,000 near-Earth asteroids, have been detected via planetary radar.\footnote{https://echo.jpl.nasa.gov/History/alldetected.html} If an ISO is close enough for such resolution, radar would be very helpful for constraining artificial profiles.

\subsection{Transmissions \& Non-Natural Emission}
\label{sec:radio}

The technosignature approaches described above rely on anomalies of the ISO compared with the properties or behavior of known minor bodies, such as kinematics or surface properties. 
Probes entering the Solar System might also produce detectable transmissions, which would be easily detectable using current facilities. We therefore should employ traditional SETI approaches to constrain technosignatures from ISOs.

Since the pioneering work of \citet{cocconi1959} and \citet{drake1961}, the primary approach for SETI has been to search for narrow-band radio emission. Radio frequencies are favorable for detection over long distances with minimal interference from the interstellar medium, and can be generated using present-day technology \citep{sheikh2025}. Despite many decades of work in this area, we are still only sensitive to a small number of needles in the cosmic haystack of detectable parameter space \citep{wright2018c}. However, since ISOs are very close during their solar system passage, the power requirements to produce a transmission detectable by our present-day radio telescopes are orders of magnitude lower than technosignature campaigns that search other stellar systems, down to the level of common household technologies (e.g., cell phones).

A range of facilities is now available for appropriate monitoring for possible radio technosignatures from ISOs. These include the purpose-built Allen Telescope Array \citep{welch2009allen}, whose primary goal is to efficiently search for technosignatures over a wide field of view. General purpose radio facilities are also available, such as the Green Bank Telescope that has been used for a variety of technosignature studies, and offers exceptional sensitivity \citep[e.g.][]{worden2017,enriquez2018}. Comets are routinely observed with radio telescopes \citep{chen2024}, and standard tracking of ISOs using non-sidereal rates should be used.

\begin{figure}[]
\centering
\includegraphics[width=3.25in]{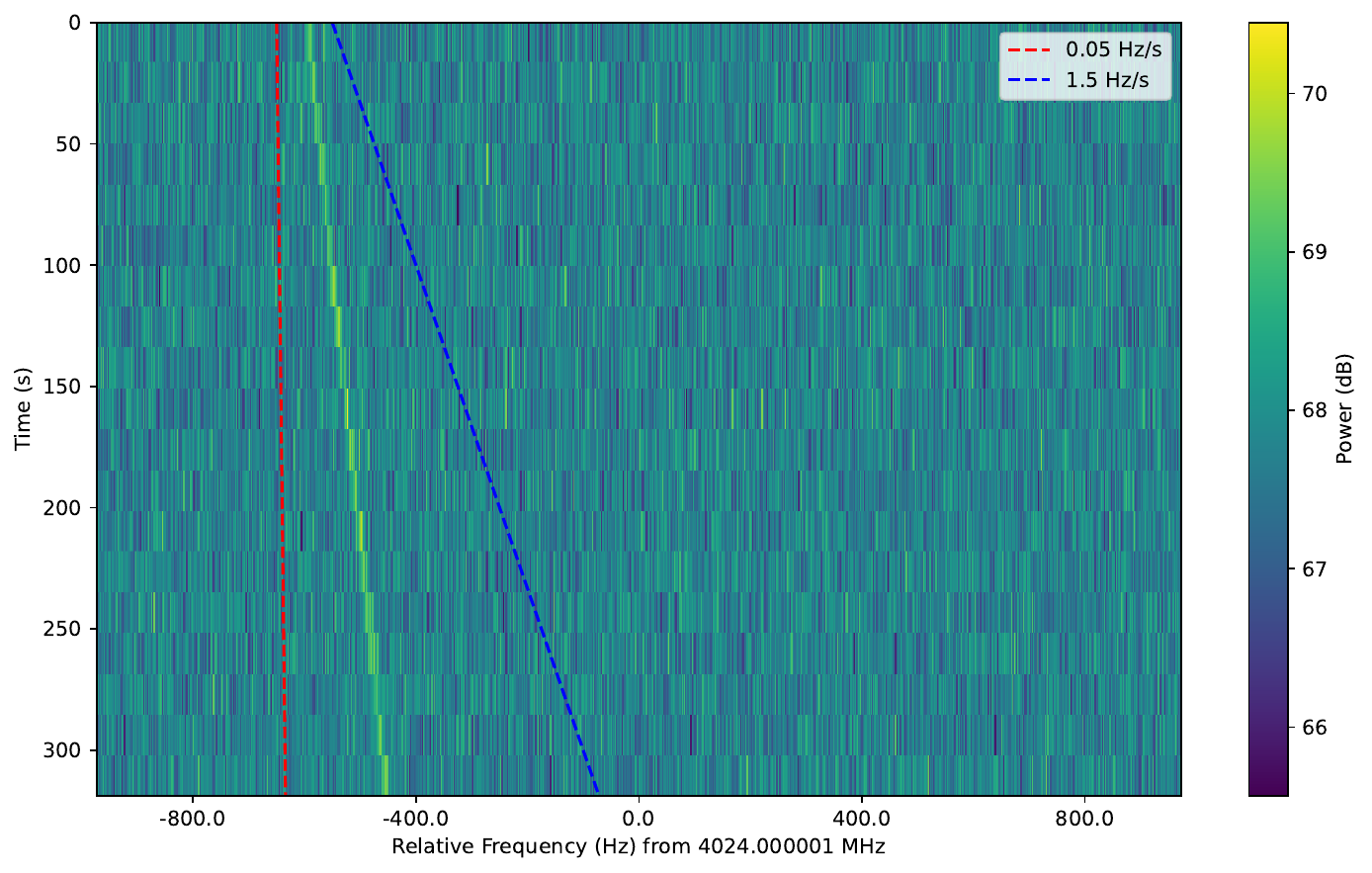}
\caption{
A simulated ``waterfall'' plot for a narrowband radio technosignature transmission from an object moving on a 3I/ATLAS-like trajectory, using \texttt{SETIgen} \citep{brzycki2022}. The measured acceleration of the ISO lets us search a specific range of drift rates (or slope in this parameter space) expected for a signal broadcast at a fixed frequency, shown as the blue and red dashed lines.
}
\label{fig:radio}
\end{figure}

In Figure \ref{fig:radio} we show a simulated narrowband radio technosignature that could be detected from an ISO like 3I/ATLAS. This ``waterfall'' plot exhibits a significant amplitude, a narrow peak that drifts in frequency over time. Since ISOs have measured accelerations as they pass through the Solar System, we can predict the drift rate (i.e., the slope in Figure \ref{fig:radio}) expected for a signal broadcast at a fixed frequency. This allows us a powerful constraint for the parameter space to search for technosignatures from ISOs, which is advantageous compared to similar studies of stars.

Frequent monitoring by such facilities for transmissions is required, particularly upon first detection of an inbound ISO (Sheikh et al. {\it in prep}), and at the times of closest approach to Earth when our sensitivity is highest, or perihelion. If a probe is instead transmitting back towards its home star system, radio monitoring during the outbound phase will be most valuable, particularly during any chance alignment of the Earth and the radiant of the ISO's origin.

There is also growing interest in detecting spectro-temporally confined electromagnetic emission at optical and IR wavelengths (e.g. lasers) as a technosignature search methodology \citep{zuckerman2023}. Lasers would appear as narrow emission lines in optical or infrared spectroscopy, and would be especially notable if they were transmitted at wavelengths where natural stellar or nebular emission lines do not occur. 
High resolution optical or infrared spectroscopy can resolve lasers over interstellar distances due to both intentional beacons, as well as leakage from e.g. light sail propulsion systems. 

Within our solar system, laser communication systems have been demonstrated to reliably achieve high-speed data transfers over many AU distances, such as the Deep Space Optical Communication (DSOC) system on the Psyche mission \citep{biswas2024}. The DSOC transmission system uses a mere 4\,W laser, transmitted with a 22-cm aperture telescope.

\begin{figure}[]
\centering
\includegraphics[width=3.25in]{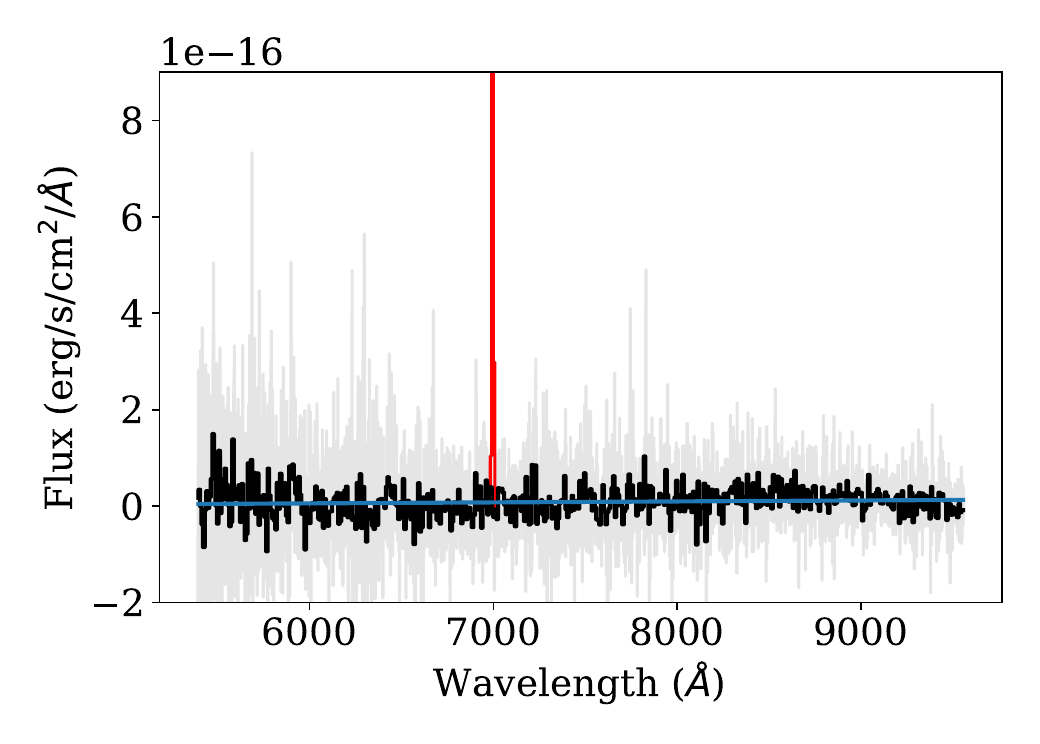}
\caption{
Simulation of a 1\,W optical laser, added to a low resolution spectrum of 3I/ATLAS from APO/KOSMOS. Even at fairly low spectral resolution, a low powered laser is extremely bright at distances of a few AU.}
\label{fig:laser}
\end{figure}

Even modestly powered lasers are easily detectable with ground-based spectroscopy. In Figure~\ref{fig:laser}, we demonstrate a simulated 1\,W optical laser at a distance of 4\,AU, observed at low resolution. We have added this laser to a real observation of 3I/ATLAS using the Apache Point Observatory (APO) 3.5-m telescope's KOSMOS spectrograph \citep{kadlec2024}. Following the calculation in \citet{zuckerman2023}, this laser has a flux amplitude of $10^{-12}\,\mathrm{erg}\,\mathrm{s}^{-1}\,\mathrm{cm}^{-2}\,\mathrm{\AA}^{-1}$. If aimed at Earth, even a low powered laser can be readily detected within the Solar System. 
As with radio SETI follow-up, optical and infrared spectra should be gathered throughout an ISO's solar system passage to constrain laser emission.

\section{Summary of Recommendations}
\label{sec:summary}

\begin{figure*}[]
\centering
\includegraphics[width=6.5in]{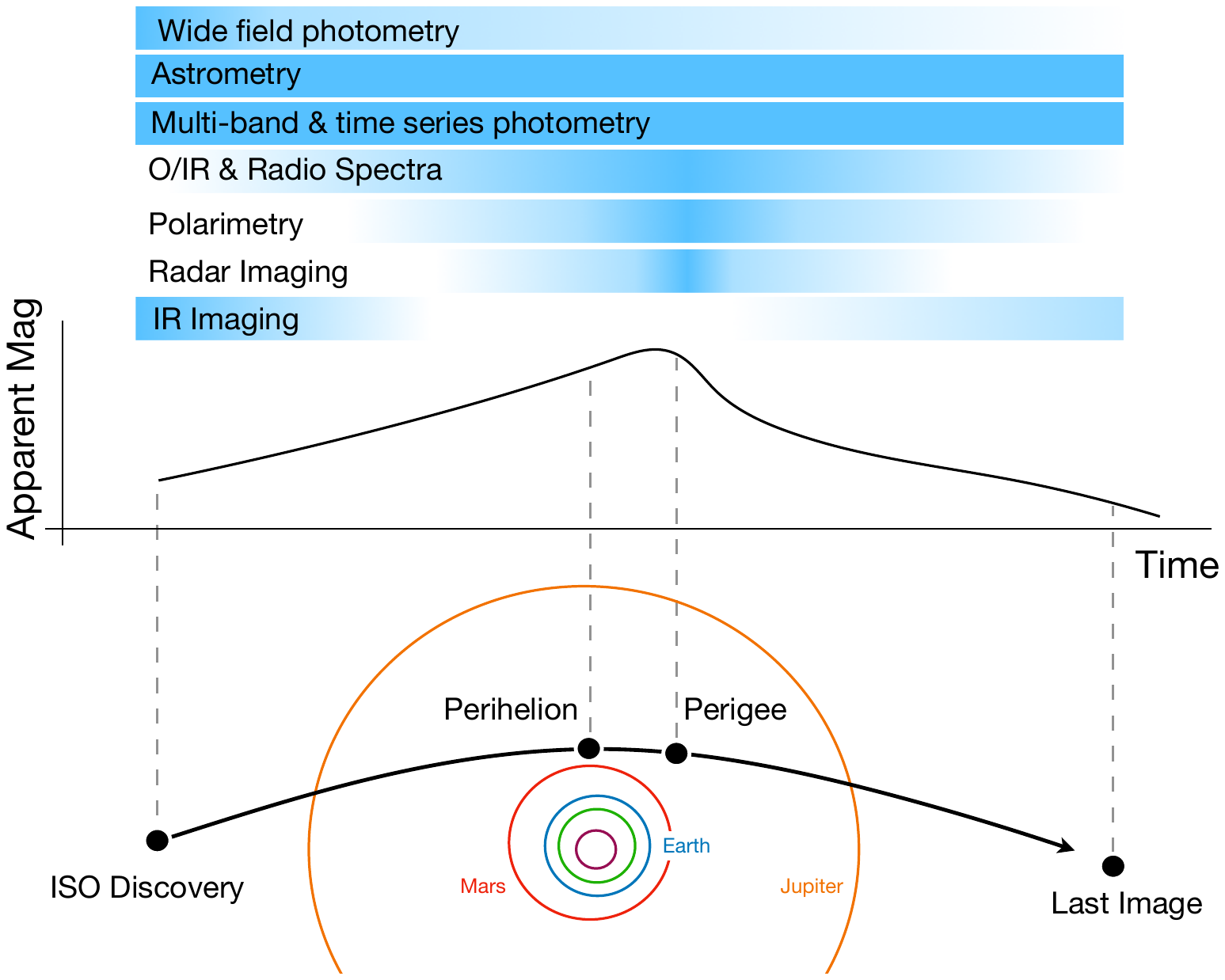}
\caption{
Schematic trajectory of an ISO as it passes through the Solar System, loosely based on the predicted orbit of 3I/ATLAS. Key phases of the orbit, including closest approach to the Sun (perihelion) and to Earth (perigee), are labeled. An illustrative light curve is shown aligned to the orbit. Observations for constraining ISO technosignatures from Table~\ref{tbl} are listed, with blue shading indicating the key times for obtaining data.
}
\label{fig:summary}
\end{figure*}

We have described in \S\ref{sec:history} and \S\ref{sec:search} a wide variety of follow-up observations that may be useful for constraining technosignatures from an ISO. These span nearly all wavelength regimes and modern observing modalities. In general, it is beneficial to observe ISOs frequently throughout their passage of the Solar System, and with multiple observing modalities, to ensure as complete a record of any transient behavior as possible.

While we have focused on the technosignature implications possible for each of these observing modes, almost all of the suggested follow-up observations are useful for other science goals relating to ISOs. For example, astrometric monitoring to detect non-gravitational accelerations will be useful in constraining the cometary nature of these objects, and such data may not need to be obtained specifically by the SETI community. As such, we recommend technosignature researchers contribute to commensal or parallel observations of ISOs, to ensure the greatest volume of data is available to the community. Similarly, technosignature studies will benefit from relying on widely accepted analysis methods for comets and ISOs, such as non-sidereal tracking and orbit modeling.

To help motivate the variety of data that is useful for constraining ISO technosignatures, we provide a summary of recommended observations in Table~\ref{tbl}. This table is not a comprehensive list of the technosignature signals possible from the data, but instead a compilation of the observing modes that should be explored. Though we include suggested timings for these observations, nearly all signals benefit from monitoring throughout the ISO's passage through the Solar System.

In Figure~\ref{fig:summary}, we present an illustrative trajectory and light curve for an ISO with an inner solar system approach. This has been loosely based on the predicted behavior of 3I/ATLAS, and the actual shape of the light curve will depend on the specifics of the ISO and Earth's orbit, as well as, for instance, cometary out-gassing. The observations for constraining technosignatures listed in Table~\ref{tbl}, with the corresponding phases of the orbit where data are recommended, are shown as shaded bars.

\begin{deluxetable*}{lll}
\tablecaption{Observing recommendations for ISO technosignatures based on the classes described in \S\ref{sec:search}. Note that the timing here is a suggestion for the most likely detection, and almost all proposed technosignatures benefit from additional data throughout the ISO passage.
\label{tbl}
}
\tablehead{
\colhead{Observing Mode} & 
\colhead{Technosignature Class} & 
\colhead{Timing}
}
\startdata
Astrometry \& tracking & Accelerations (\S3.1) & Full arc of passage \\
Full sky photometry & Early detection \& origin (\S3.1) & Earliest possible \\
Optical/IR spectroscopy & Spectral anomalies, lasers (\S3.2, 3.4)  & Closest approach \\
Radio spectroscopy & Transmissions (\S3.4)  & Closest approach \\
Multi-band photometry & Color \& phase curve anomalies (\S3.2, 3.3)   & Full arc of passage \\
High-cadence imaging & Rotational modulation (\S3.3)   & Multi-night campaign \\
Infrared photometry & Waste heat (\S3.2)  & Pre- and post-perihelion \\
Polarimetry & Unusual surface properties (\S3.2)  & Closest approach \\
Radar imaging  & Shape anomalies (\S3.3)  &  When within radar range \\
\enddata
\end{deluxetable*}

\section{Discussion}
\label{sec:dicussion}

ISOs represent an especially important target for technosignature searches. While detection of ISOs in wide field surveys is quite recent, their unique potential for SETI has been long established. Literature from the past 60 years has shown that our own solar system is one of the most favorable domains for carrying out statistically complete SETI campaigns, and objects with likely interstellar origins are particularly noteworthy for such investigation. 

The coming generation of wide field survey telescopes will naturally produce a steady stream of such targets, and the SETI community should prepare to participate in rapid follow-up and monitoring campaigns. As technosignatures are possibly one of the most unambiguous and longest lived signals for the detection of life \citep{wright2022}, these observations allow facilities like the Rubin Observatory to play a critical role in astrobiology over the coming decade \citep{davenport2019a, parts2025}.

During the final preparation of our manuscript, \citet{eldadi2025} proposed a scale to quantify the significance of anomalies detected for ISOs, based on the Torino Impact Hazard Scale defined for near-Earth asteroids \citep{torino}. This is conceptually similar to the ``Rio Scale'' to assess technosignature significance more broadly \citep{rio1,rio2}, with subjective focus on comparing ISOs to solar system objects. As \citet{eldadi2025} indicate with their ``Loeb Scale'', and as we have outlined in detail, anomalies from technosignatures could arise across many axes. A broad search for many forms of outliers is therefore critical for advancing SETI within the Solar System.
However, as discussed throughout our review of the possible technosignatures that can be detected, probabilistic assessment based on comparisons to solar system bodies is limited by our lack of constraint on the underlying properties of the ISO population. To date, no credible evidence for technosignature signals has been found from the three known ISOs. 
While work has begun on statistical treatments for these new and fascinating objects \citep[e.g.][]{Marceta2023a,hopkins2025}, and a small number of nearly unambiguous technosignatures may exist, we believe an ISO-specific scale for technosignature assessment is premature.

Any potential detection of technosignatures from an ISO will require the most stringent and detailed confirmation possible. This includes validation of the raw data, analysis methodology, and if possible independent observation of the signal \citep{sheikh2021}. Since plausible technosignatures range from the subtle to the unambiguous \citep{sheikh2020}, the significance of the signal and any possible natural contamination must be clearly defined \citep[e.g. see][]{rio2,eldadi2025}. Guidelines for validation and community notification of signals exit, such as those defined by the IAA SETI Committee Post-Detection Protocols \footnote{https://iaaseti.org/en/declaration-principles-concerning-activities-following-detection/}, and should be followed to ensure both the safety of scientists and rigor of the results.

While we have focused on the technosignature opportunities possible from monitoring of ISOs throughout their solar system passage, it is worth noting that a wide variety of ancillary science can be advanced from such observations. Studies of surface composition, comet activity and acceleration, shape, and rotation all require identical or highly complementary observations to those needed for numerous technosignature studies. As 3I/ATLAS continues to be observed, for example, we encourage teams to collaborate and share their observations (both raw and reduced data) to enable a variety of scientific goals.

Several of the proposed scenarios for interstellar probes in the literature highlight the possibility of technology being connected to natural objects such as asteroids or comets (e.g. buried in, or sitting on, the surface). These natural objects may have even been modified in a number of ways, such as hollowed-out asteroids with stable rotation to generate an interior surface and spin gravity for habitation \citep{Miklavcivc2022}. In addition, since we can detect very low power communication systems from objects within a few AU, it is important to continue observing ISOs throughout their passage, and with many wavelengths and facilities. 
The presence of ``normal'' behavior, such as natural cometary activity, or surface colors consistent with bare asteroids, should not deter our follow-up observations aimed at constraining technosignatures. 
As with all technosignature searches, if we only look once, we may simply miss an incredibly obvious transmission or signal.

\begin{acknowledgments}
The authors thank Mario Jurić, Meg Schwamb, Pedro Bernardinelli, and Jon Giorgini for helpful discussions.

J.{}R.{}A.{}D.\ and C.{}O.{}C.\ acknowledge support from the DiRAC Institute in the Department of Astronomy at the University of Washington. The DiRAC Institute is supported through generous gifts from the Charles and Lisa Simonyi Fund for Arts and Sciences, Janet and Lloyd Frink, and the Washington Research Foundation.

The authors acknowledge support from Breakthrough Listen. The Breakthrough Prize Foundation (\url{https://breakthroughprize.org/}) funds the Breakthrough Initiatives, which manage Breakthrough Listen.

The Center for Exoplanets and Habitable Worlds and the Penn State Extraterrestrial Intelligence Center are supported by Penn State and its Eberly College of Science.

E.{}W.\ is supported by the National Science Foundation
Graduate Research Fellowship Program under Grant
No.~1000383199. Any opinions, findings, and conclusions or recommendations expressed in this material are those of the author(s) and do not necessarily reflect the views of the National Science Foundation.

E.{}E.{}Y.\ was funded as a participant in the Berkeley SETI Research Center Research Experience for Undergraduates Site, supported by the National Science Foundation under Grant No.~2244242.

D.{}Z.{}S.\ is supported by an NSF Astronomy and Astrophysics Postdoctoral Fellowship under award AST-2303553. This research award is partially funded by a generous gift of Charles Simonyi to the NSF Division of Astronomical Sciences. The award is made in recognition of significant contributions to Rubin Observatory’s Legacy Survey of Space and Time. 

A.{}F.\ and S.{}S.\ gratefully acknowledge funding via NASA Exobiology program under grant 80NSSC20K0622.

T.{}J.{}W.{}L.\ acknowledges the ideas and advice from the participants in the ``Data-Driven Approaches to Searches for the Technosignatures of Advanced Civilizations'' workshop organized by the W.~M.~Keck Institute for Space Studies.
Part of this research was carried out at the Jet Propulsion Laboratory, California Institute of Technology, under a contract with the National Aeronautics and Space Administration.

This research has made use of NASA’s Astrophysics Data System.

Based on observations obtained with the Apache Point Observatory 3.5-meter telescope, which is owned and operated by the Astrophysical Research Consortium.

\end{acknowledgments}

\bibliography{references,refs}

\end{document}